\documentclass[aps,preprint,showpacs]{revtex4}
\usepackage{graphicx}
\newcommand{\e}{\textnormal{e}}

\begin{document}

\date{\today)}

\title[Quantum noise]{Quantum noise in optical interferometers}

\author{Volodymyr G. Voronov}
\affiliation{Faculty of Physics, Taras Shevchenko National
University of Kyiv, Kyiv, Ukraine}
\author{Michael Weyrauch}
\affiliation{Physikalisch-Technische Bundesanstalt, D-38116
Braunschweig, Germany}

\begin{abstract}
We study the photon counting noise in optical interferometers used
for gravitational wave detection. In order to reduce quantum noise a
squeezed vacuum is injected into the usually unused input port. It
is investigated under which conditions the gravitational wave signal
may be amplified without increasing counting noise concurrently.
Such a possibility was suggested as a consequence of the
entanglement of the two output ports of a beam splitter. We find
that amplification without concurrent increase of noise is not
possible for reasonable squeezing parameters.  Photon distributions
for various beam splitter angles and squeezing parameters are
calculated.

\end{abstract}

\maketitle

\section{Introduction}\label{sec:introduction}

Optical interferometers for the detection of gravitational waves
need an extremely accurate control of various sources of noise. In
an effort to reduce quantum-mechanical noise in such
interferometers, Caves proposed the squeezed state
technique~\cite{caves:1}: into the normally unused port of the
interferometer a squeezed vacuum state is injected. Details of this
technique have since been analyzed in many experimental and
theoretical investigations (e.g. Refs.~\cite{yurke:1, assaf:1}).

A central element of an interferometer is a beam splitter, and as
can be easily shown, a photon state, which is a product of the
states in each input port, is not a product of the photon states in
each output port of the beam splitter. Instead, the two output
states of a beam splitter are entangled, and it was the subject of
recent work by Barak and Ben-Aryeh~\cite{barak:1} to investigate the
consequences of this entanglement for the photon statistics of an
optical interferometer.

In particular, it was suggested that under certain conditions the
gravitational wave signal may by amplified significantly without a
corresponding increase in counting noise. This effect was attributed
to the entanglement effects mentioned above in connection with
squeezing.

It is the purpose of this paper to investigate this surprising
proposal in detail. To this end we determine the photon
distributions in the output state of a beam splitter for weak and
strong coherent states injected into one of the input ports of the
beam splitter in addition to the squeezed vacuum in the other. Our
results are in disagreement with Ref.~\cite{barak:1} for both weak
and strong input states. In particular, we cannot confirm the
finding in Ref.~\cite{barak:1} that the gravitational wave signal
may be amplified without concurrent increase of noise.

In order to handle the entanglement effects one needs to disentangle
exponential operators. This is rather simple in the present case,
and we describe a Lie algebraic method, where the disentangling
coefficients are calculated numerically from a set of coupled
nonlinear equations. We provide a complete graphical overview of the
disentangling coefficients as a function of the squeezing factor of
the injected squeezed vacuum state and the angle of the beam
splitter.

In section 2 we develop a general formula for the calculation of the
photon number distributions in the two output ports of a beam
splitter. Furthermore, we discuss the special case where the beam
splitter is oriented close to 90$^\circ$ to an incoming strong
coherent state with a squeezed vacuum entering the other port.
Numerical results will be discussed in section 3. A brief summary
concludes the paper. Technical details are relegated to a few
Appendices.

\section{Photon statistics in a Michelson interferometer}\label{sec:approximation}

We consider a beam splitter and inject a coherent state into port 1
and a squeezed vacuum state into port 2. The incoming photon state
is therefore described by
\begin{equation}
|\psi_{\rm in}(\alpha,\zeta)\rangle = \hat{S}_2(\zeta)\hat{D}_1(\alpha)|0,0\rangle
\end{equation}
with
\begin{equation}
\hat{D}_1(\alpha)=\exp\left(\alpha \hat{a}_1^\dagger -\alpha^*\hat{a}_1\right),\;\;\;
\hat{S}_2(\zeta)=\exp\left(\frac{\zeta^*}{2} \hat{a}_2^{ 2} -\frac{\zeta}{2} \hat{a}_{2}^{\dagger2}\right).
\end{equation}
The coherence parameter $\alpha$ and the squeezing parameter $\zeta$
are complex numbers.  The photon field operators $\hat{a}_i$ and
$\hat{a}_j^\dagger$ fulfill the boson commutation relation
$[\hat{a}_i,\hat{a}_j^\dagger]=\delta_{i,j}$.

After passing the beam splitter the field is described by the
rotated field operators $\hat{b}_i$ and $\hat{b}_i^\dagger$ with
$[\hat{b}_i,\hat{b}_j^\dagger]=\delta_{i,j}$~\cite{campos:1}
\begin{equation}
\left(
  \begin{array}{c}
    \hat{a}_1 \\
    \hat{a}_2 \\
  \end{array}
\right)=
\left(
  \begin{array}{cc}
    \cos\gamma & \sin\gamma \\
    -\sin\gamma & \cos\gamma \\
  \end{array}
\right)
\left(
  \begin{array}{c}
    \hat{b}_1 \\
    \hat{b}_2 \\
  \end{array}
\right).
\end{equation}
The parameter $\gamma$ parameterizes the splitting ratio of the beam
splitter with respect to the incoming beam in port 1.

Now expressing the $\hat{a}$ operators in terms of the $\hat{b}$
operators enables us to write the photon state after passing the
beam splitter as follows
\begin{equation}\label{eq:output}
|\psi_{\rm out}(\alpha,\zeta,\gamma)\rangle = \exp(|\zeta|\hat{A}) \hat{D}_1(\alpha\cos\gamma) \hat{D}_2(\alpha\sin\gamma) |0,0\rangle
\end{equation}
with
\begin{equation}\label{eq:A}
\hat{A}=\hat{s}_1 \sin^2\gamma+\hat{s}_2\cos^2\gamma+\hat{s}_{12}\sin\gamma\cos\gamma
\end{equation}
and
\begin{equation}\label{eq:oper1}
\hat{s}_i=\frac{1}{2|\zeta|} (\zeta^* \hat{b}_i^{2}-\zeta \hat{b}_i^{\dagger2})\;\;\;\;\;\;\;
\hat{s}_{12}=\frac{1}{|\zeta|}(\zeta \hat{b}_1^\dagger \hat{b}_2^\dagger
     -\zeta^* \hat{b}_1 \hat{b}_2 ).
\end{equation}
In the absence of the operator $\hat{s}_{12}$ the output would be
obviously characterized by squeezed coherent states in both output
ports. However, these states are entangled via $\hat{s}_{12}$ and
the output state cannot be factorized. This fact significantly
complicates evaluation of the photon statistics of the output state.

However, it is possible to use Lie algebraic disentangling
techniques in order to rewrite the output state in a way which
enables the determination of photon distributions. To this end we
first define the additional operator $\hat{t}_{12}$,
\begin{equation}\label{eq:oper2}
\hat{t}_{12}= \hat{b}_1 \hat{b}_2^\dagger- \hat{b}_1^\dagger \hat{b}_2.
\end{equation}
The operators $\hat{s}_1, \hat{s}_2, \hat{s}_{12}$, and
$\hat{t}_{12}$ form a closed Lie algebra with the commutation
relations,
\begin{eqnarray}\label{eq:commutators}
& &[\hat{s}_1,\hat{s}_2]= 0,\quad  [\hat{s}_{12},\hat{t}_{12}]=-2\hat{s}_1+2\hat{s}_2,\nonumber\\
& &[\hat{s}_1,\hat{s}_{12}]=\hat{t}_{12},\quad[\hat{s}_2,\hat{s}_{12}]=-\hat{t}_{12},\\
& &[\hat{s}_1,\hat{t}_{12}]=\hat{s}_{12},\quad [\hat{s}_2,\hat{t}_{12}]=-\hat{s}_{12}.\nonumber
\end{eqnarray}
As a consequence it is possible to write Eq.~(\ref{eq:output}) as
follows~\cite{wei:1, scholz:2}
\begin{equation}\label{eq:output}
|\psi_{\rm out}\rangle = \exp(\sigma_{T} \hat{t}_{12})\exp(\sigma_{S} \hat{s}_{12})\exp(\sigma_1\hat{s}_1)\exp(\sigma_2\hat{s}_2)
\hat{D}_1(\alpha \cos\gamma) \hat{D}_2(\alpha\sin\gamma)
|0,0\rangle.
\end{equation}
The output state is now expressed in terms of two squeezed coherent
states entangled via the operators $\exp(\sigma_{T} \hat{t}_{12})$
and $\exp(\sigma_{S} \hat{s}_{12})$.  The coefficients $\sigma_{T}$,
$\sigma_{S}$, $\sigma_1$, $\sigma_2$ are real functions of the input
parameters $r=|\zeta|$ and $\gamma$. A simple method for the
numerical determination of these parameters is described in
Appendix~\ref{se:disentangling}.

In order to determine the photon statistics of the output state we
need to determine its number (Fock) representation. We start out
from the number representation of a squeezed coherent
state~\cite{scully:1}
\begin{equation}\label{eq:focksqch}
\hat{S}(\zeta) \hat{D}(\beta)|0\rangle=
\sum_{n=0}^\infty \frac{1}{\sqrt{n!}}f_{n}(\zeta,\beta)  |n\rangle
\end{equation}
with
\begin{equation}
f_{n}(\zeta,\beta)=\frac{(\e^{i\theta}\tanh r)^{n/2}}{2^{n/2} (\cosh r)^{1/2}}
\exp\left(-\frac{1}{2}(|\beta|^2-\e^{-i\theta}\beta^2 \tanh r)\right) H_n\left(\frac{\beta \e^{-i\theta/2}}
     {\sqrt{2\cosh r \sinh r}}\right),
\end{equation}
$\zeta=r \e^{i\theta}$, and $H_n$ the Hermite polynomials. Now
applying the two exponential operators $\exp(\sigma_{T}
\hat{t}_{12})$ and $\exp(\sigma_{S} \hat{S}_{12})$ on a product of
such states yields the Fock representation of the output state.
After some straightforward but tedious algebra one obtains the
photon distribution $P_{n_1n_2}$ in the two output ports of the
interferometer
\begin{equation}\label{eq:phodi}
P_{n_1n_2}(\alpha,\zeta,\gamma)=\frac{n_2!}{n_1!}\left|\sum_{m_4=0}^{n_2}\sum_{m_3=0}^{n_1+m_4}\sum_{m_2=0}^{M_2}\sum_{m_1=0}^\infty
A_{n_1n_2}^{m_1m_2m_3m_4}f_{N_1+m_1}(\e^{i\theta}\sigma_1,\beta_1)f_{N_2+m_1}(\e^{i\theta}\sigma_2,\beta_2)\right|^2
\end{equation}
with
\begin{eqnarray}
& & N_1=n_1-m_2-m_3+m_4, \quad N_2=n_2-m_2+m_3-m_4,\nonumber\\
& &M_2=\min(n_1-m_3+m_4, n_2+m_3-m_4),\\
& &A_{n_1n_2}^{m_1m_2m_3m_4}=\frac{\nu_S^{m_1} \lambda_S^{m_2}\nu_T^{m_3}\lambda_T^{m_4}}{m_1!m_2!m_3!m_4!}\frac{
(n_1+m_4)!}{(n_2-m_4)!}\frac{(n_2+m_3-m_4)!}{N_1!N_2!}\times\nonumber\\
& &\quad\quad\quad\quad\quad\quad\quad \e^{\mu_S (1+n_1+n_2-2 m_2)/2}\e^{\mu_T (-n_1+n_2-2m_4)/2},\nonumber\\
& & \zeta=r \e^{i\theta},\quad \beta_1=\alpha \cos\gamma, \quad \beta_2=\alpha\sin\gamma,\nonumber\\
& & \lambda_S= \e^{i\theta}\tanh\sigma_S, \quad \mu_S=-2\log(\cosh \sigma_S)),\quad    \nu_S=\e^{-i\theta}\tanh\sigma_S,\nonumber\\
& & \lambda_T= \tan\sigma_T,  \quad \mu_T=-2\log(\cos \sigma_T)), \quad    \nu_T=\tan\sigma_T.\nonumber
\end{eqnarray}
The sum over $m_1$ in Eq.~(\ref{eq:phodi}) is in principle
unrestricted above. In practice a suitable upper limit must be
chosen such that the probability is correctly normalized. Most of
the numerical results in the following are calculated from
Eq.~(\ref{eq:phodi}). A few details used for the derivation of
Eq.~(\ref{eq:phodi}) are given in Appendix~\ref{se:eval}.

In addition to the general case discussed above, we investigate the
special case $\gamma=\pi/2+\delta$ for which one finds (see
Appendix~\ref{se:disentangling}) $\sigma_1=r$, $\sigma_2=0$,
$\sigma_S=-\delta\sinh r$, and $\sigma_T=\delta(1-\cosh r)$, if
$\delta$ is sufficiently small. We furthermore assume a very strong
coherent state incoming in port 1, such that the $b_2$ and
$b_2^\dagger$ operators can be replaced in the entanglement factors
in Eq.~(\ref{eq:output}) by their expectation values $\alpha$ and
$\alpha^*$, respectively. The output state can then be written as
\begin{equation}\label{eq:app_out}
|\psi_{\rm out}\rangle=\hat{D}_1(-\alpha\delta(1-\cosh r))\hat{D}_1(-\delta\alpha^* \e^{i\theta}\sinh r)
\hat{S}_1(\zeta)\hat{D}_1(-\alpha \delta)\hat{D}_2(\alpha)|0,0\rangle
\end{equation}
with $\zeta=r \e^{i\theta}$. A similar result was obtained along
somewhat different lines in Ref.~\cite{barak:1}, however with
opposite sign in the first factor. The operators with index 1 may be
combined using the relations given in
Appendix~\ref{se:transformations} and we obtain for the output state
\begin{equation}\label{eq:outpihalf}
|\psi_{\rm out}\rangle=\e^{i |\alpha|^2 \delta^2\Delta}\hat{S}_1(\zeta)\hat{D}_1(-\alpha\delta\kappa)\hat{D}_2(\alpha)|0,0\rangle
\end{equation}
with $\alpha= |\alpha| \e^{i\phi}$ and
\begin{eqnarray}\label{eq:kappa}
\kappa(r,\theta-2\phi)&=&|\kappa|\e^{i \lambda}=\cosh r+\e^{i (\theta-2\phi)} \sinh r, \\
\Delta(r,\theta,\phi)&=&-\frac{1}{2} \sin(\theta - 2\phi) \e^{-2i \phi}\sinh(2 r)\nonumber.
\end{eqnarray}
As one can see from Eq.~(\ref{eq:outpihalf}), a strong coherent
state with coherence parameter $\alpha$ exits through port 2 of the
interferometer and a  weak squeezed coherent state with coherence
parameter $-\alpha\delta\kappa$ and squeezing parameter $\zeta$
exits through port 1. Note, that the coherence parameter depends on
the squeezing parameter. The phase factor $\e^{i|\alpha|^2
\delta^2\Delta}$ is irrelevant for the determination of the photon
statistics.

The photon statistics in port 1 is immediately obtained from
Eq.~(\ref{eq:focksqch})
\begin{equation}\label{eq:darkdist}
P_{n_1}=\frac{1}{n_1!}\left|f_{n_1}(\zeta,-\alpha\delta\kappa)\right|^2.
\end{equation}
The mean and the variance of this distribution may be obtained
analytically~\cite{scully:1}
\begin{eqnarray}\label{eq:meanvari}
\langle n_1 \rangle & =&\delta^2|\alpha|^2(\cosh 2r + \cos(\theta-2\phi)\sinh 2r)\times\nonumber\\
 & &\quad(\cosh 2r - \cos(\theta-2\phi-2\lambda))\sinh 2r) + \sinh^2 r,\nonumber\\
 \langle\Delta n_1^2 \rangle &=& \delta^2|\alpha|^2 (\cosh 2r +\cos(\theta-2\phi)\sinh 2r)\times\\
     & & \quad(\cosh 4r-\cos(\theta-2\phi-2\lambda))\sinh 4r) +2 \sinh^2 r \cosh^2 r.\nonumber
\end{eqnarray}
The phase $\lambda(r,\theta-2\phi)$ of $\kappa$ defined in
Eq.~(\ref{eq:kappa}) depends on the squeezing factor $r$ and the
angle $\theta-2\phi$. This dependence complicates the interpretation
of these formulas. We will discuss results calculated from
Eqs.~(\ref{eq:darkdist}) and (\ref{eq:meanvari}) in
section~\ref{se:strong}.

\section{Numerical results}\label{se:numres}

\begin{figure}
\unitlength1cm
\begin{picture}(18,13)(0,0)
 \put(0,0)  {\includegraphics[width=6.5cm]{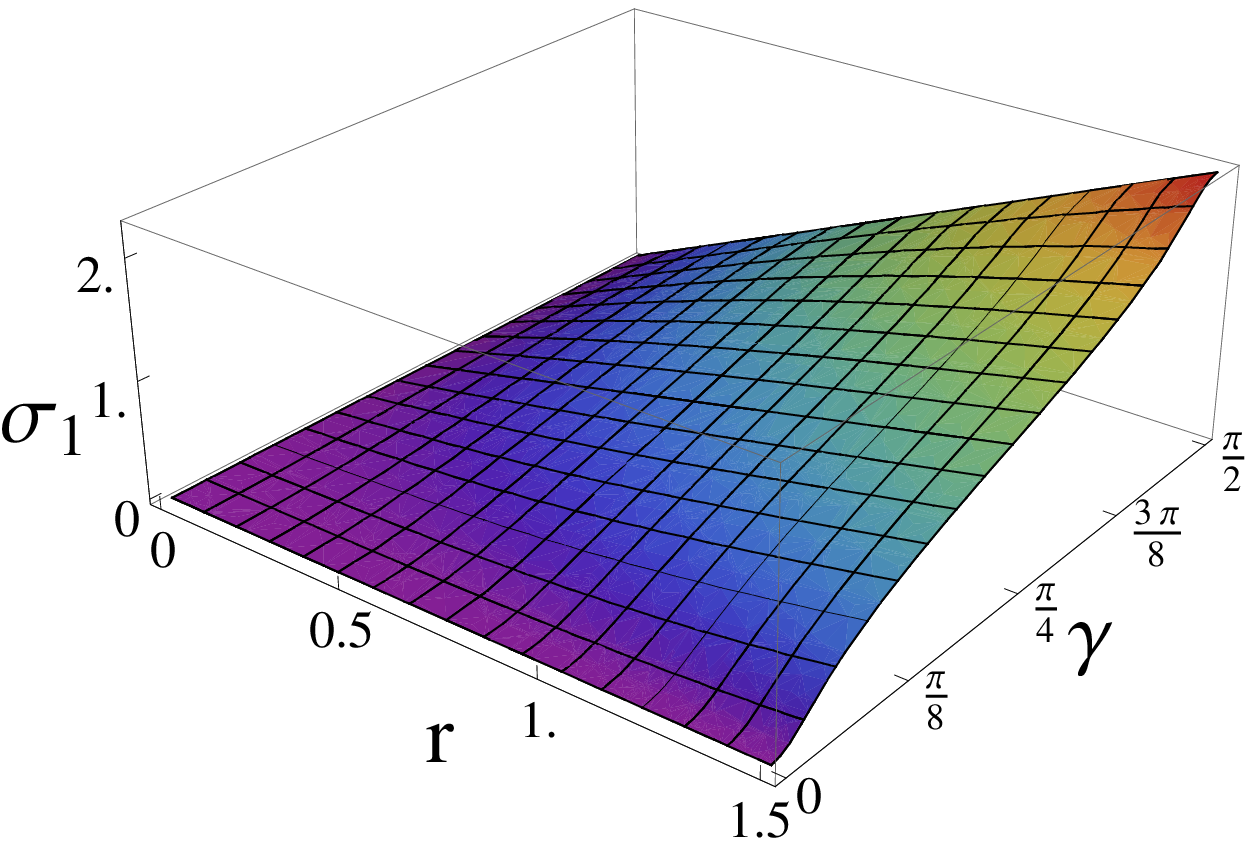}}
 \put(7.5,0.){\includegraphics[width=6.5cm]{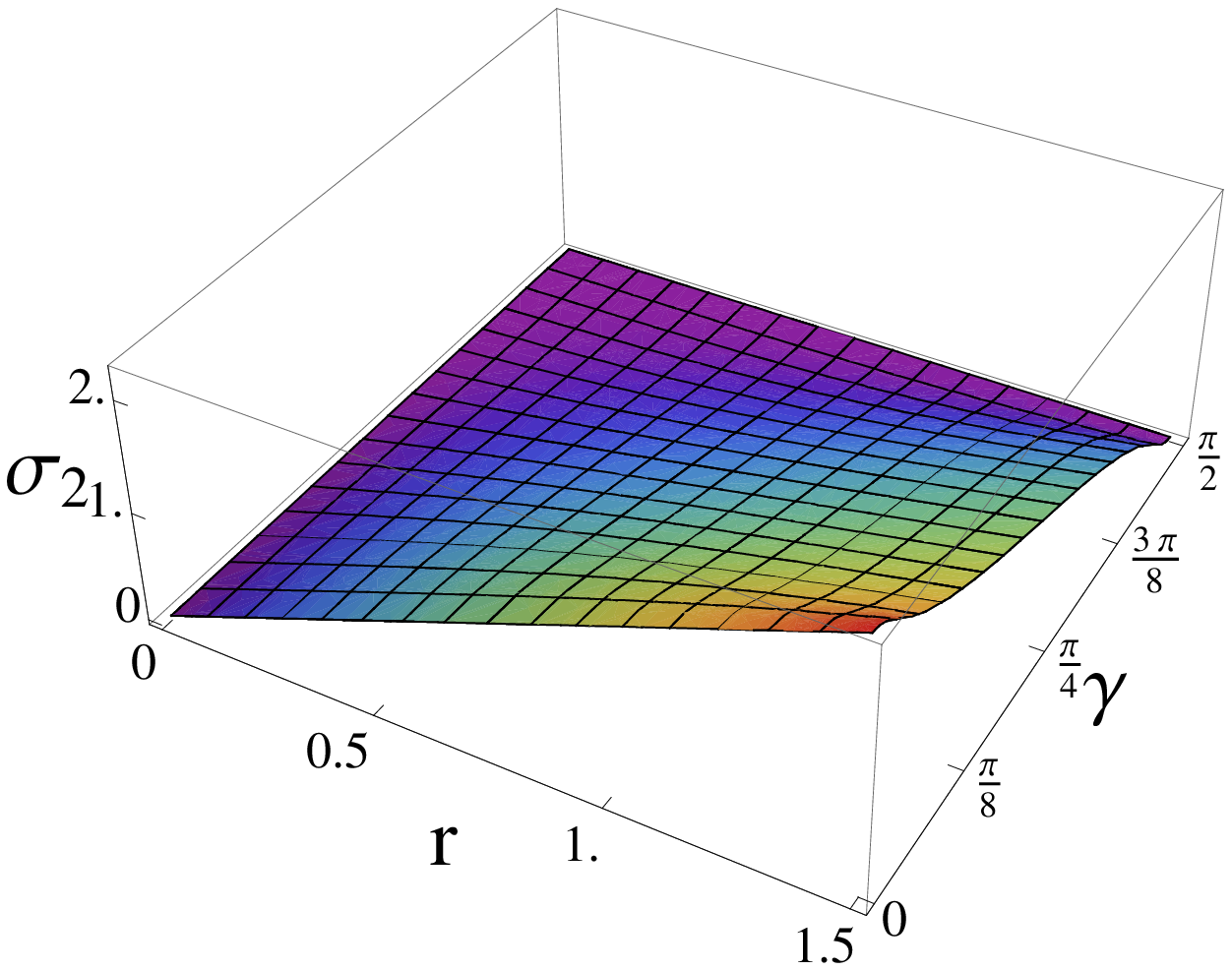}}
 \put(0.,6.5)  {\includegraphics[width=6.5cm]{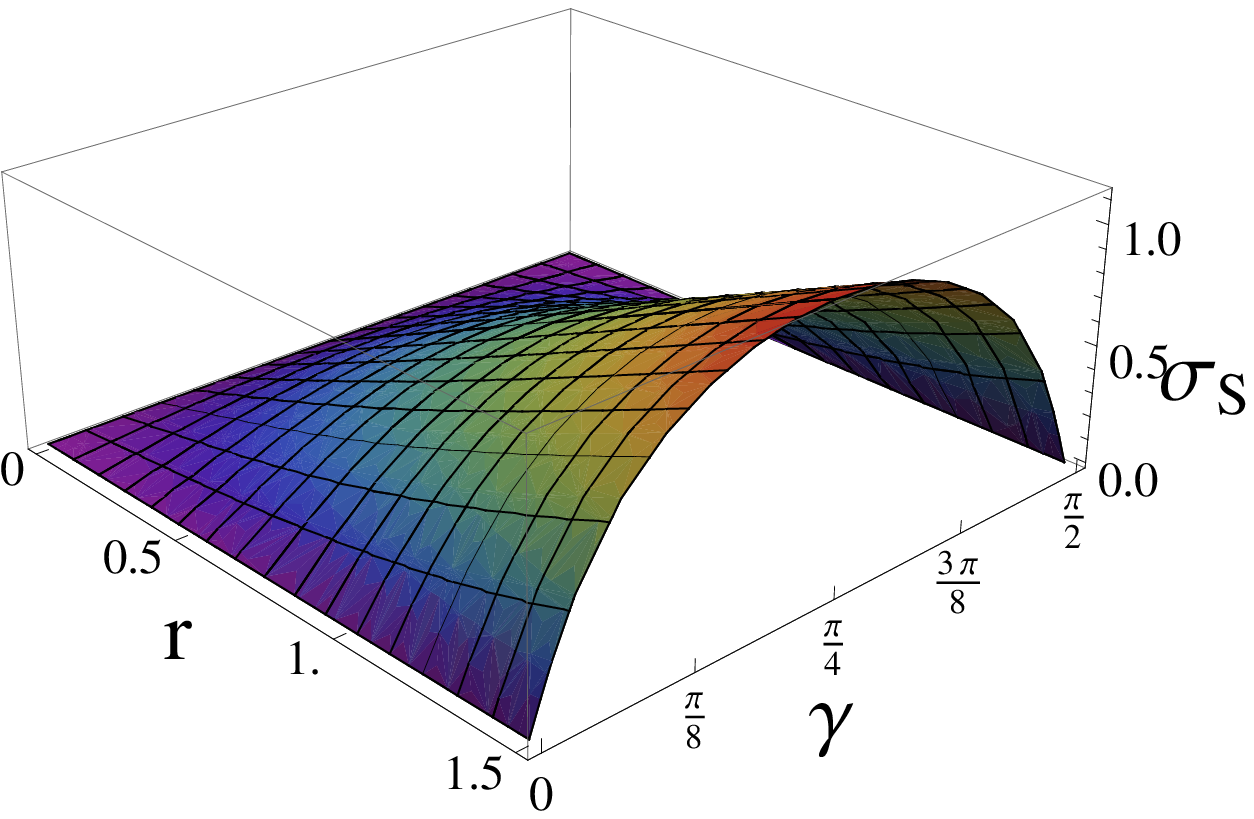}}
 \put(7.5,6.5){\includegraphics[width=6.5cm]{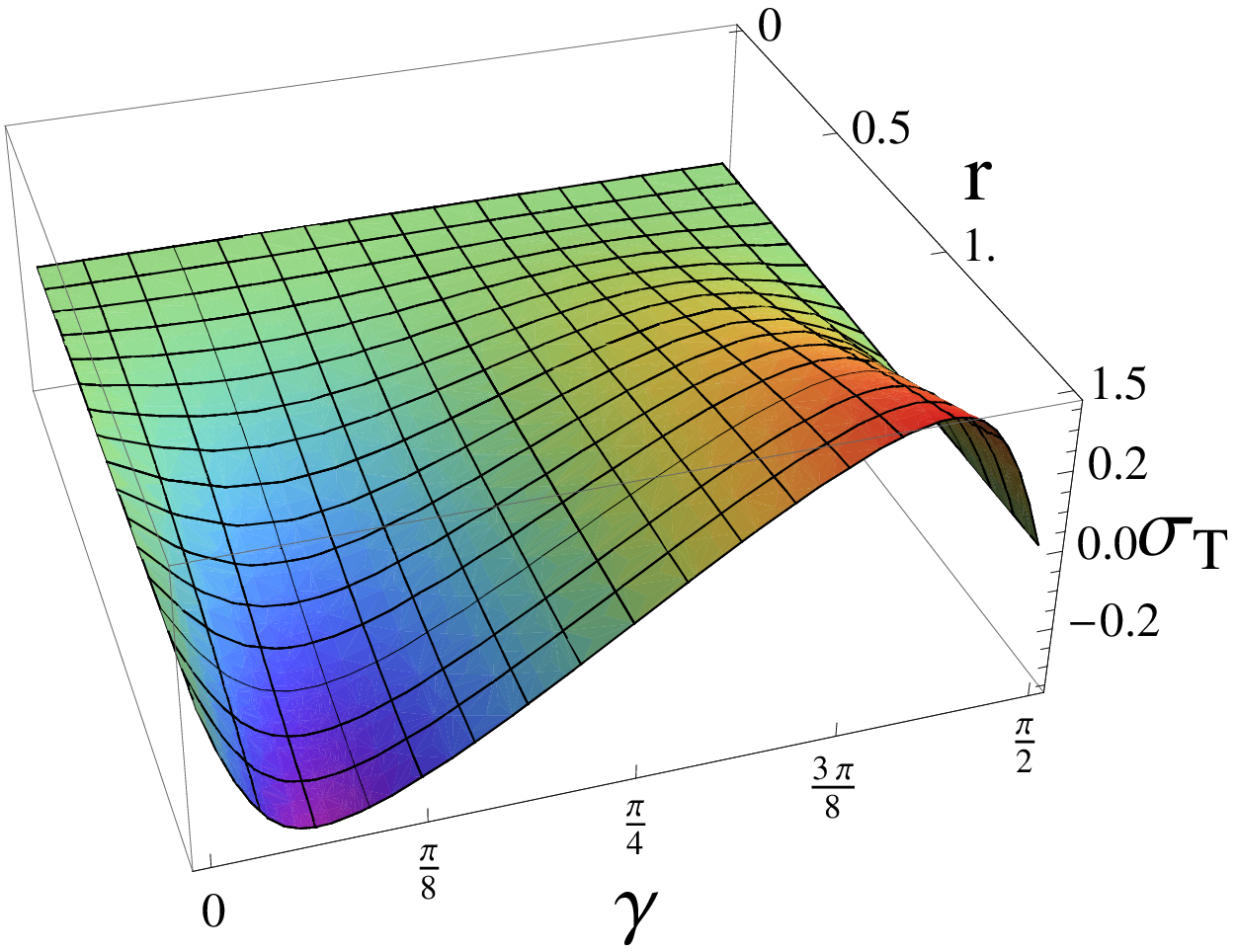}}
\end{picture}
\caption{(Color online) Disentangling coefficients $\sigma_1$, $\sigma_2$, $\sigma_S$, and $\sigma_T$ as a function of
the beam splitter angle $\gamma$ and squeezing factor $r$.}\label{fig:discoeff}
\end{figure}

As is obvious from Eqs.~(\ref{eq:output}) and (\ref{eq:phodi}), in
order to determine the photon statistics of the output state one
needs to calculate the disentangling coefficients  $\sigma_1$,
$\sigma_2$, $\sigma_S$, and $\sigma_T$. They are easily obtained as
solutions of the nonlinear equations~(\ref{eq:nonlinear}) derived in
Appendix~\ref{se:disentangling}. Numerical results are plotted in
Fig.~\ref{fig:discoeff}.
The figures clearly show the special values $\sigma_1= r$,
$\sigma_2=0$ , $\sigma_S=0$ , and $\sigma_T=0$ for $\gamma=\pi/2$,
and  $\sigma_1=r/2$, $\sigma_2=r/2$, $\sigma_S=r/2$, and
$\sigma_T=0$ for $\gamma=\pi/4$. We consider squeezing factors up to
$r=1.5$ in our numerical work, since at the present time the largest
squeezing factor experimentally realized is about $r=1.3$
corresponding to a maximum squeezing of about -11.5
dB~\cite{mehmet:1}.

\subsection{Weak coherent field and $\gamma=\pi/4$ and $\gamma=\pi/8$}\label{se:weak}

\begin{figure}
\unitlength1cm
\begin{picture}(18,7)(0,0)
\put(0,0)      {\includegraphics[width=10.cm]{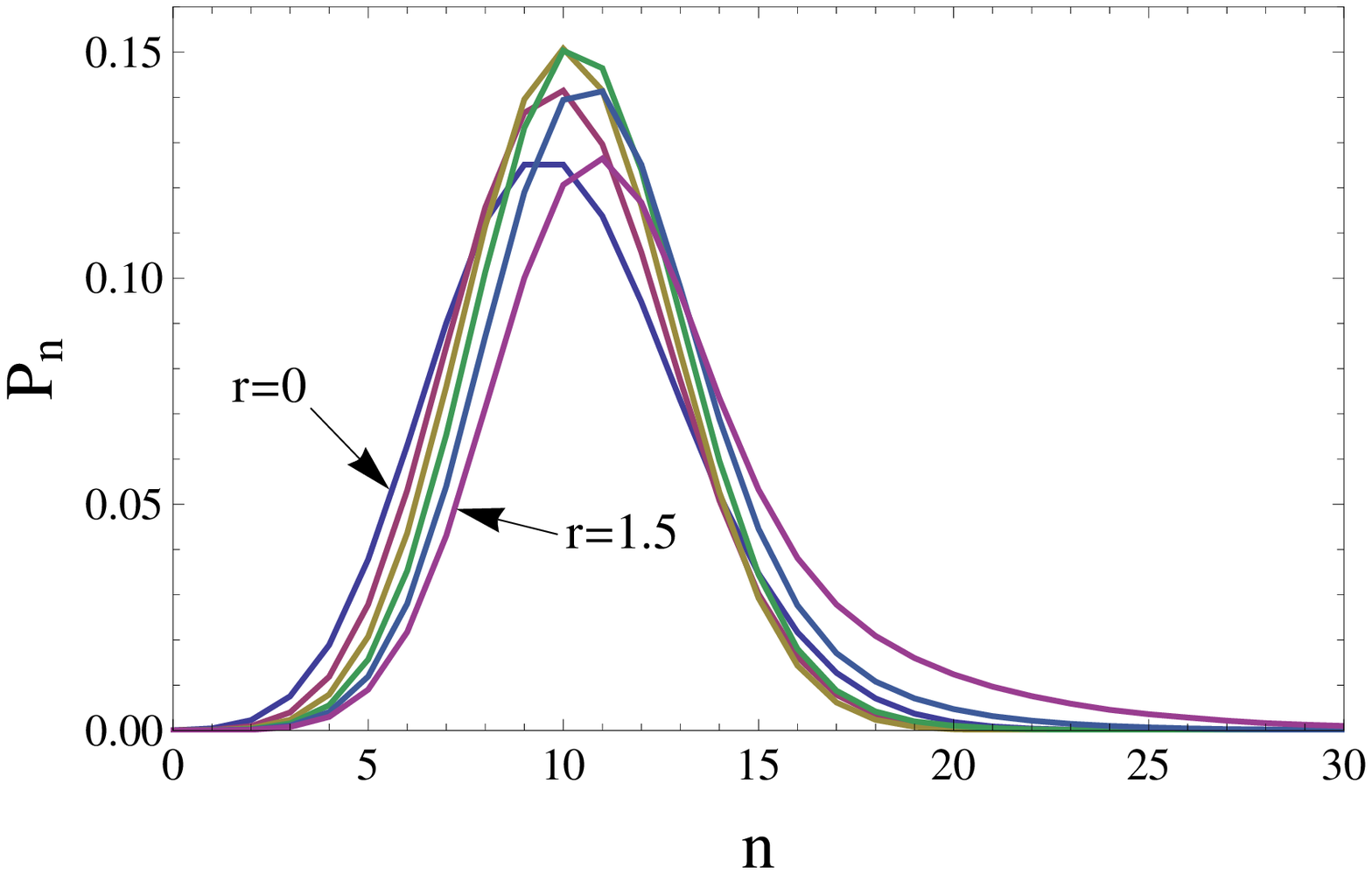}}
\put(10.1,3.5) {\includegraphics[width=4. cm]{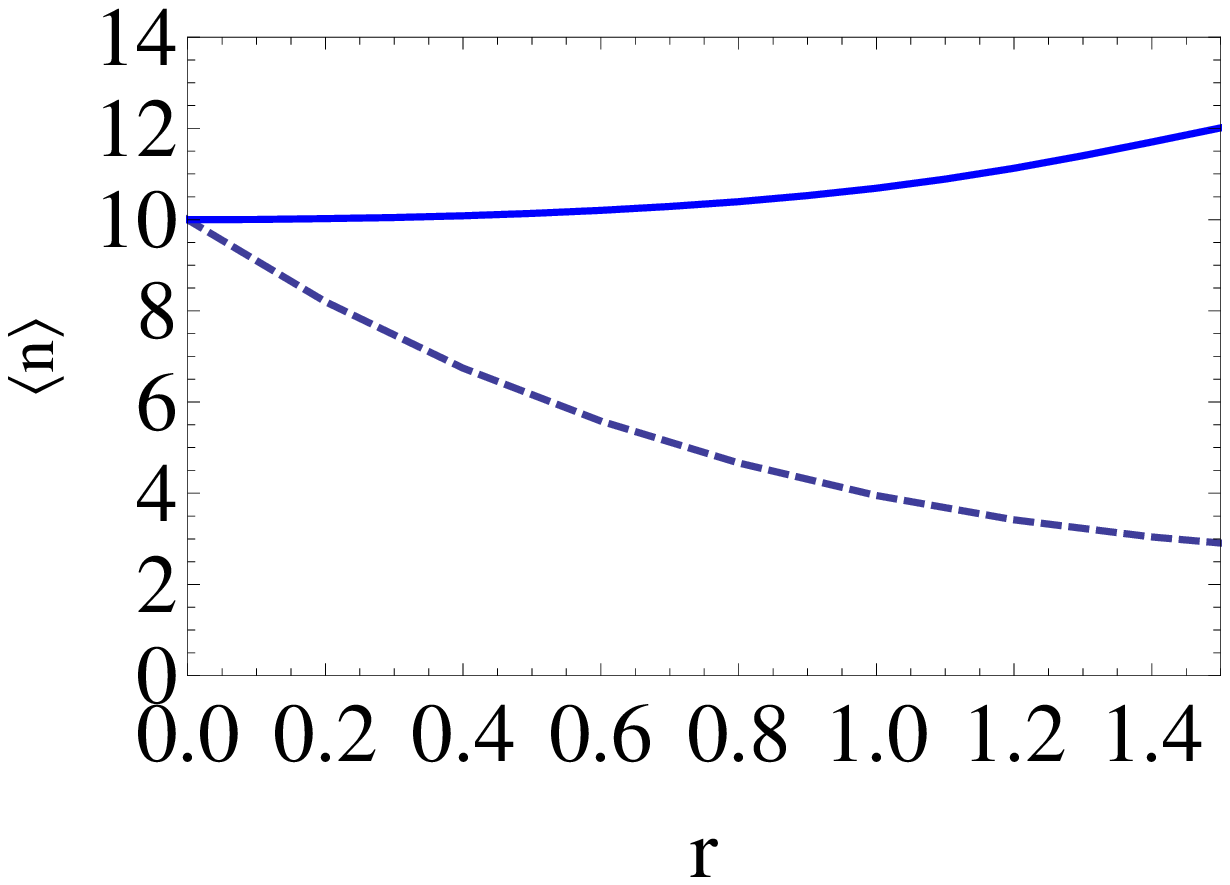}}
\put(10.0,.5)  {\includegraphics[width=4. cm]{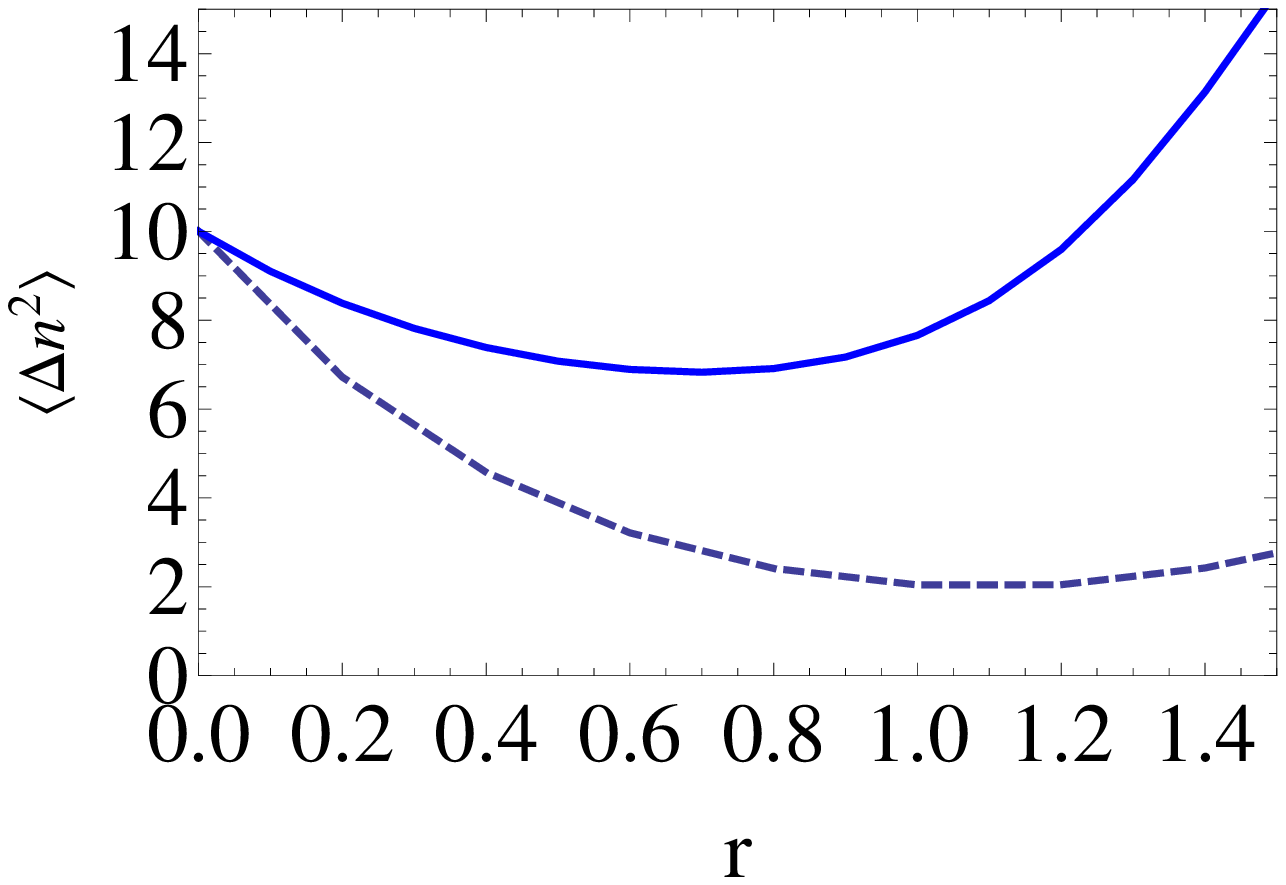}}
\end{picture}
\caption{(Color online) Photon distributions for $|\alpha|^2=20$, $\gamma=\pi/4$,
and various squeezing factors $r=0,~0.3,~0.6,~0.9,~1.2,~1.5$. On the right of the figure we plot the mean and the
variance squared of the distributions as a function of the squeezing factor $r$ (full lines). The dashed lines
show calculations without the entangling factors.}\label{fig:pi4}
\end{figure}
In Fig.~\ref{fig:pi4} we plot photon distributions in the output
port 1 of a beam splitter calculated from Eq.~(\ref{eq:phodi}) for a
weak coherent field ($|\alpha|^2=20$) injected into input port 1.
The beam splitter is set at an angle $\gamma=\pi/4$, and squeezed
vacua with various squeezing parameters $r$ are injected into port
2. For $\gamma=\pi/4$ one finds analytically $\sigma_1=r/2,\quad
\sigma_2=r/2,\quad \sigma_S=r/2$, and $\sigma_T=0$, and as a
consequence, the entanglement operator $\e^{\sigma_T \hat{t}_{12}}$
equals unity. We observe that squeezing slightly reduces the
variance of the distributions in the output port up to about
$r=0.7$, and that it slightly shifts the mean of the distributions
to larger photon numbers. This particular case was studied before by
Barak and Ben-Aryeh~\cite{barak:1}, and we note that our results are
in contrast to Ref.~\cite{barak:1} where a down shift of the mean
was predicted. (We also cannot confirm that the probabilities
calculated in Ref.~\cite{barak:1} are correctly normalized.) We
checked, that the reasons for the discrepancies between our results
and Ref.~\cite{barak:1} are missing faculty symbols in all factors
under the square roots in equations (36) and (39) of
Ref.~\cite{barak:1}. They arise if one applies a power of a creation
or destruction operator on a Fock state.

The results shown Fig.~\ref{fig:pi4} clearly indicate that in order
to determine the photon statistics of the output of a beam splitter
the entanglement factors must not be neglected. In fact,
calculations without the entanglement factors would predict an
improvement of the counting statistics by injecting a squeezed
vacuum, which is not justified. Nevertheless there is an optimum
squeezing parameter of about $r=0.7$, where we find a small
improvement of the counting statistics by squeezing.

\begin{figure}
\unitlength1cm
\begin{picture}(18,7)(0,0)
\put(0,0) {\includegraphics[width=10.cm]{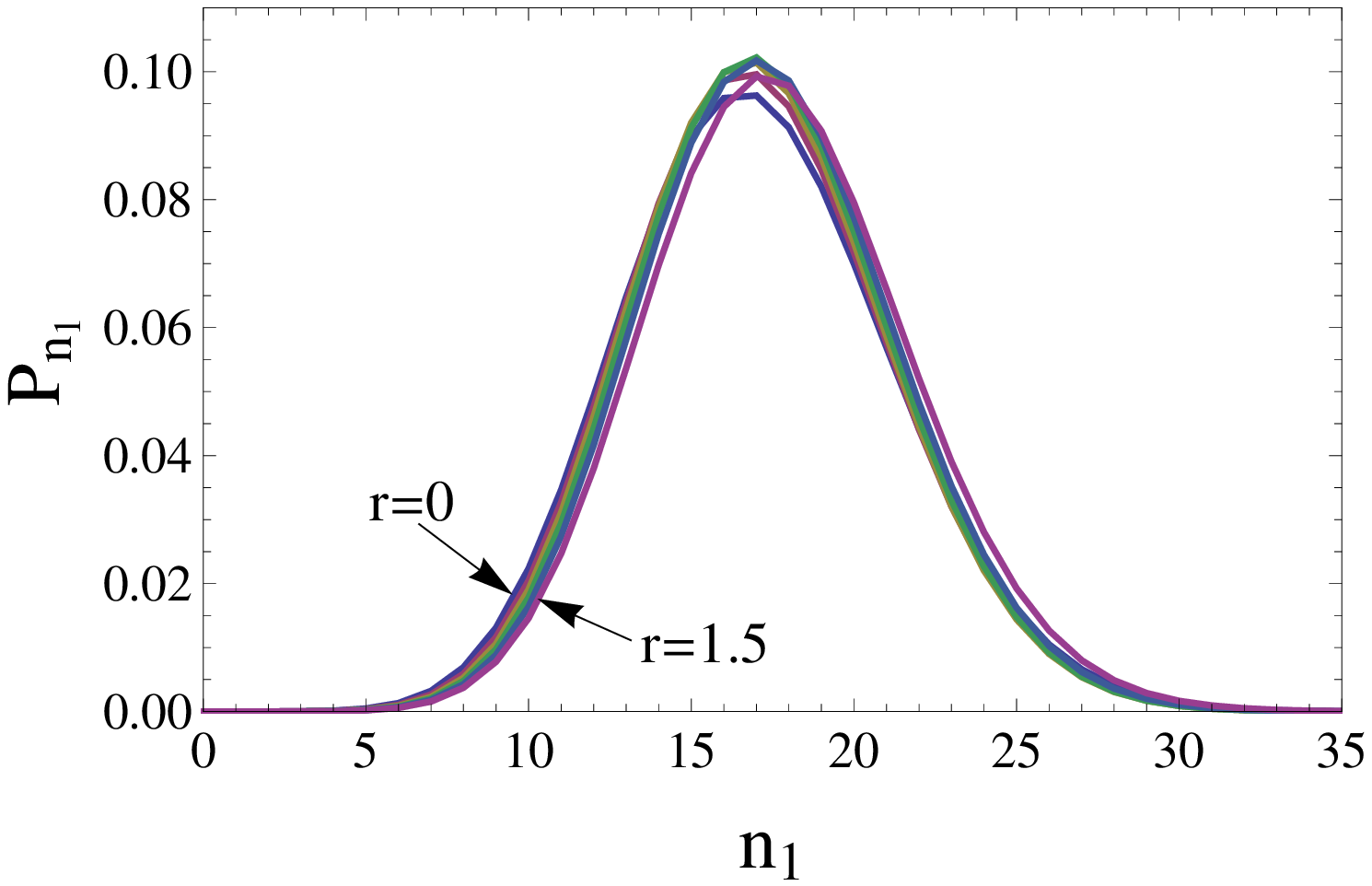}}
\put(10.1,3.5){\includegraphics[width=4. cm]{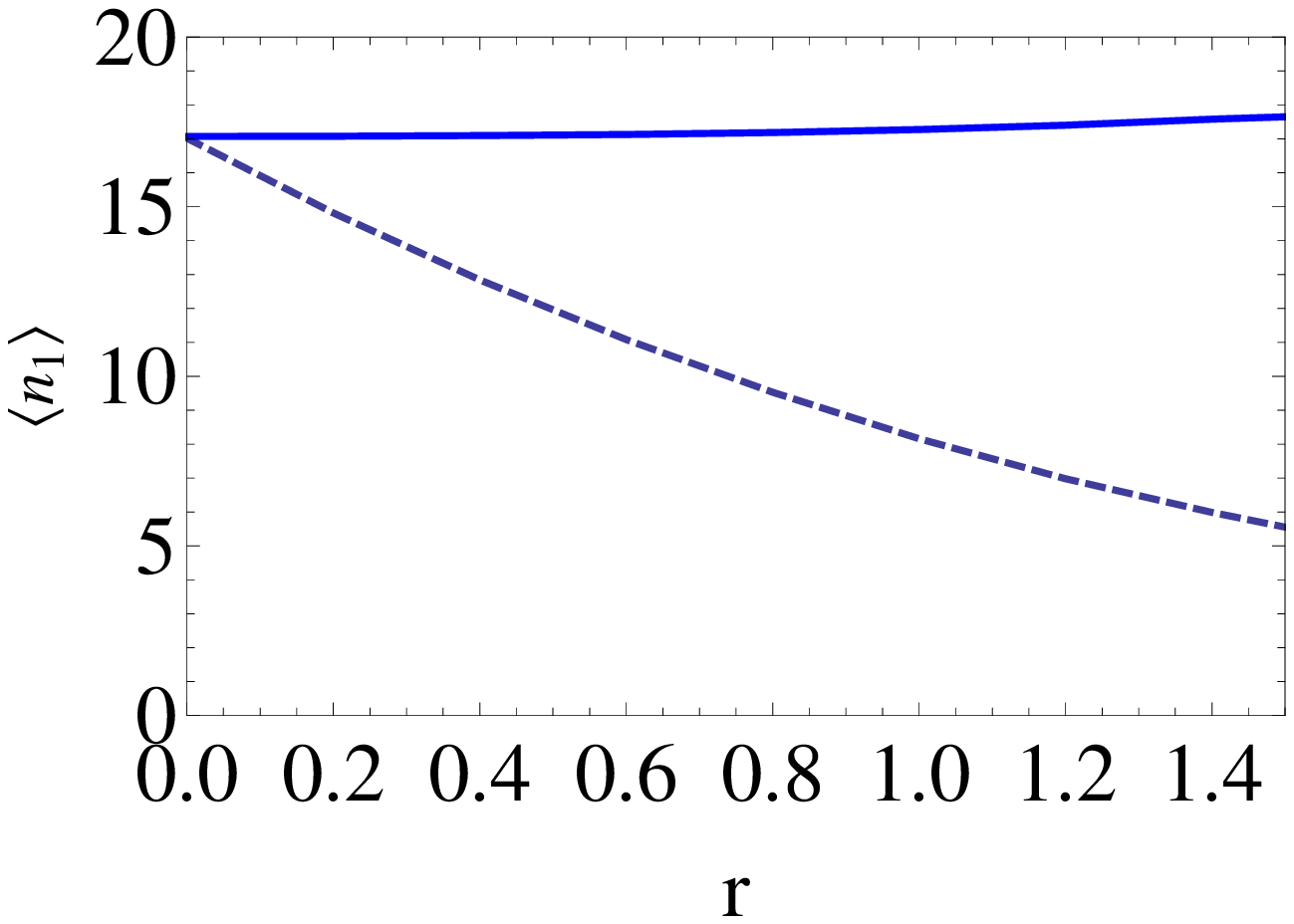}}
\put(10.0,.5){ \includegraphics[width=4. cm]{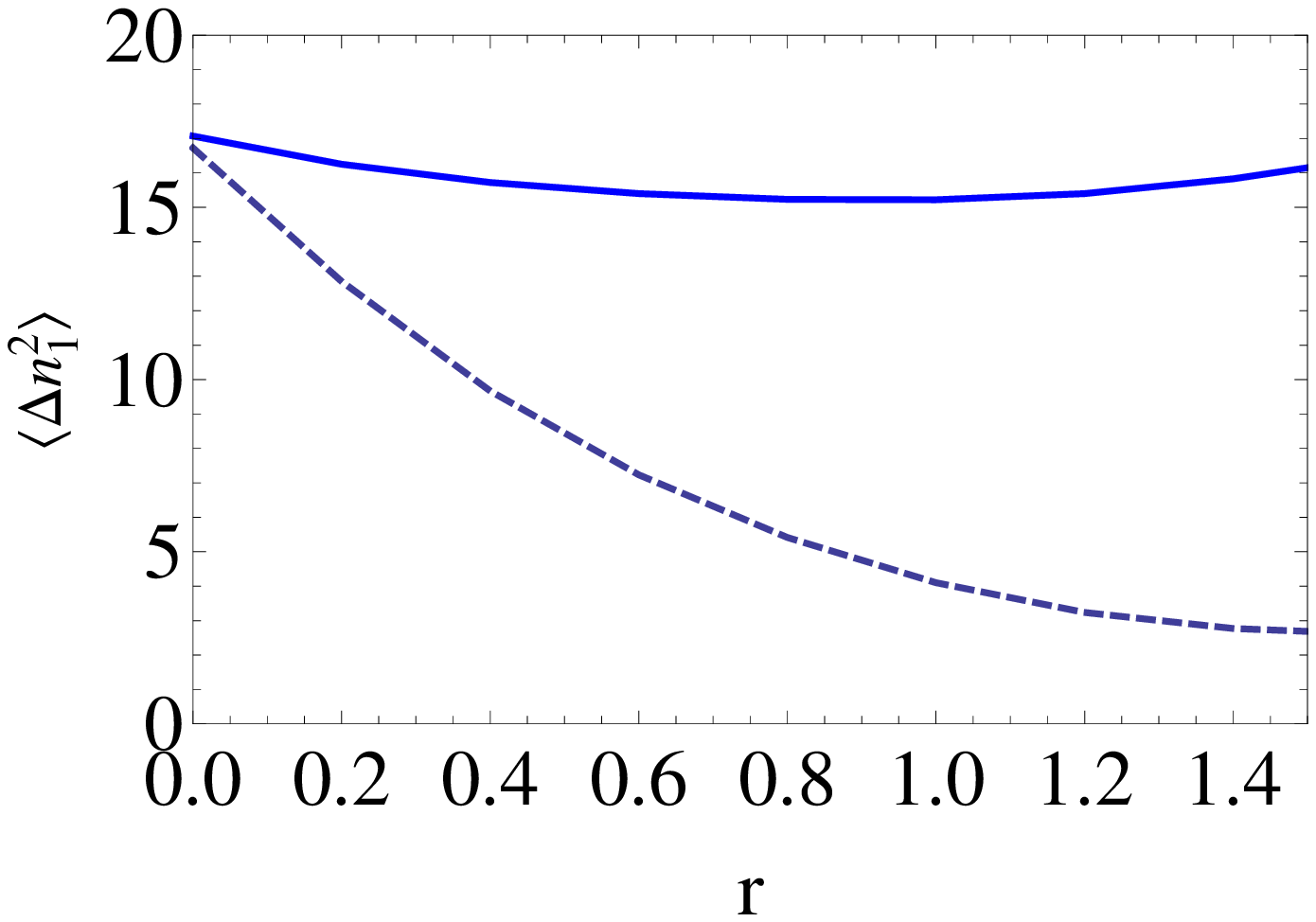}}
\end{picture}
\caption{(Color online) Photon distributions for $|\alpha|^2=20$, $\gamma=\pi/8$,
and various squeezing factors $r=0,~0.3,~0.6,~0.9,~1.2,~1.5$. On the right of the figure we plot the mean and the
variance squared of the distributions as a function of the squeezing factor $r$ (full lines). The dashed lines
show calculations without the entangling factors.}
\label{fig:pi8}
\end{figure}

Fig.~\ref{fig:pi8} shows similar results for a beam splitter set at
an angle $\gamma=\pi/8$ and otherwise the same parameters. For this
angle we need the full numerical solution for the disentangling
coefficients as discussed above, in particular, in this case both
entangling operators $\e^{\sigma_T \hat{t}_{12}}$ and $\e^{\sigma_S
\hat{s}_{12}}$ contribute. Here, we see  squeezing essentially has
no effect on the distributions. Both the mean and the variance
cannot be significantly influenced by squeezing.
Figure~\ref{fig:pi8} shows that calculations without the
entanglement factors produce a completely wrong indication of an
improved photon statistics gained by injecting a squeezed vacuum.

\subsection{Strong coherent field and $\gamma=\pi/2\pm\delta$}\label{se:strong}

\begin{figure}
\unitlength1cm
\begin{picture}(18,5)(0,0)
 \put(0,0)  {\includegraphics[width=6.5cm]{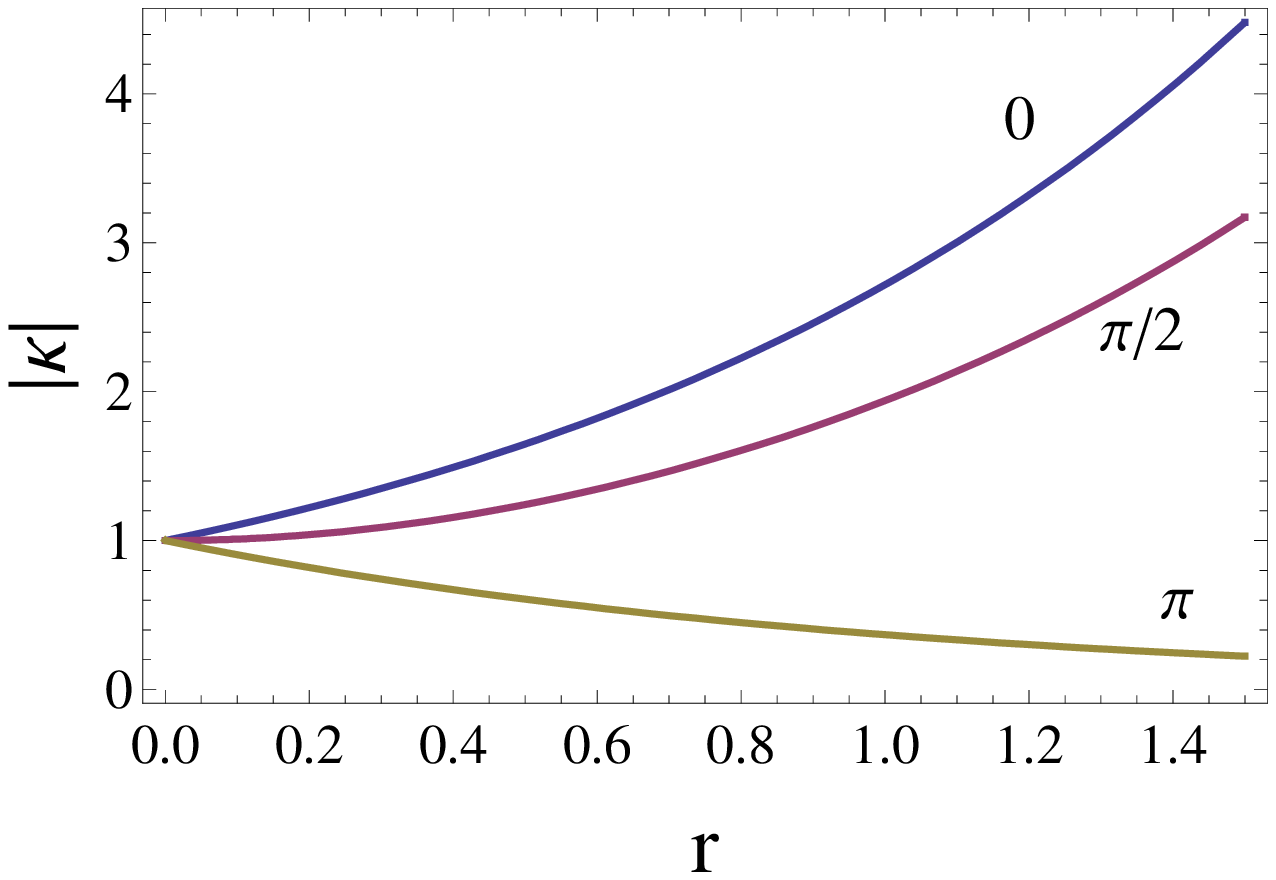}}
 \put(7.5,0.){\includegraphics[width=6.5cm]{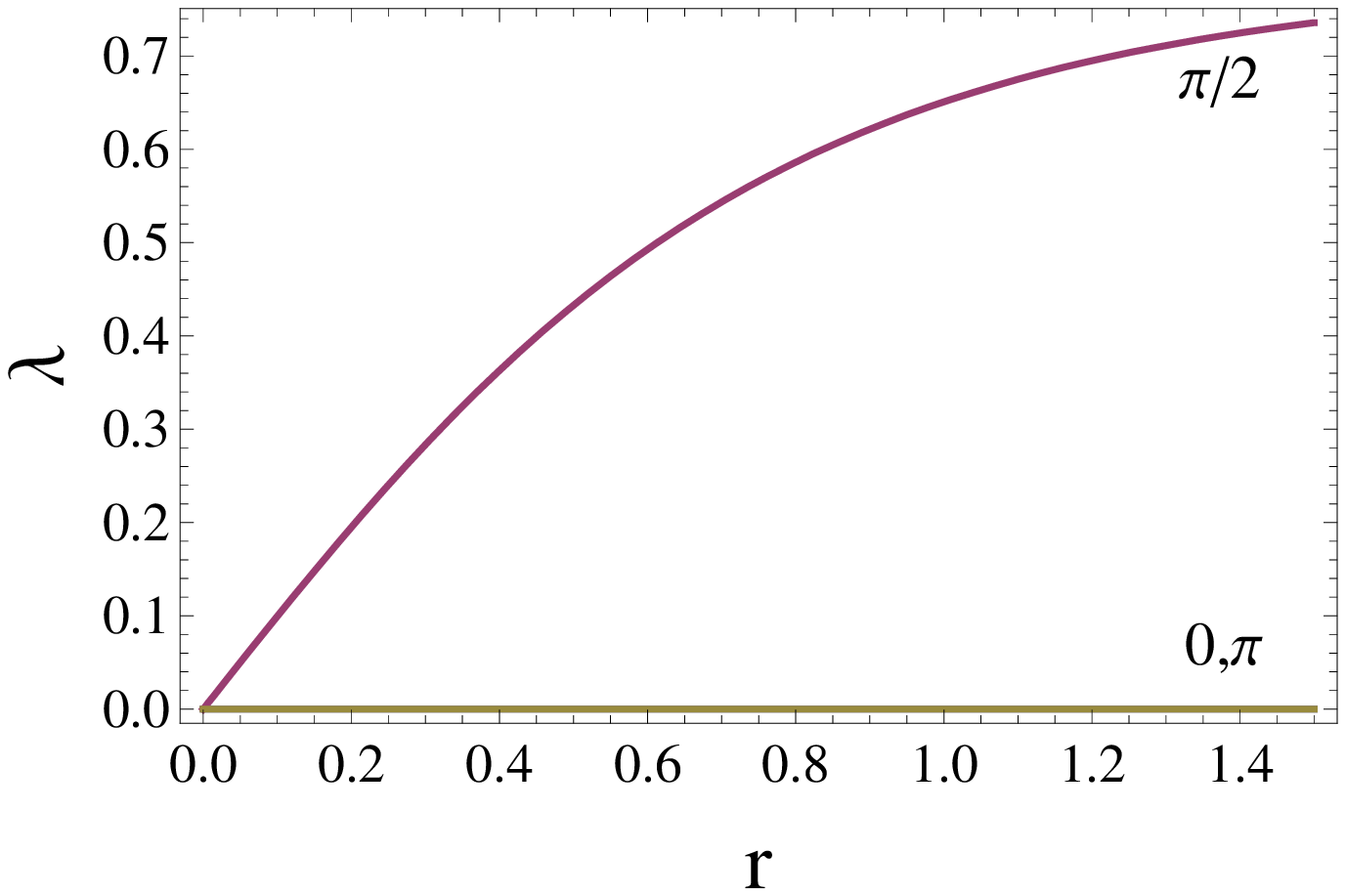}}
\end{picture}
\caption{(Color online) Modulus $|\kappa|$ and phase $\lambda$ of the coherence
parameter $\kappa=|\kappa|\e^{i\lambda}=\cosh r+\e^{i (\theta-2\phi)} \sinh r$ as a function of the
squeezing factor $r$. The label of the curves indicates the angle $\theta-2\phi$.}\label{fig:kappa}
\end{figure}

In our second numerical example, we study a strong coherent field
injected into port 1 of the beam splitter. According to
Eq.~(\ref{eq:app_out}) the output in port 2 is then a strong
coherent state, and the output in port 1 (dark port) a weak squeezed
coherent state. Its coherence parameter $-\alpha\delta\kappa(\zeta)$
depends on the squeezing parameter $\zeta$ of the squeezed vacuum
state injected into input port 2. In the Fig.~\ref{fig:kappa} the
dependence of $\kappa$ on the squeezing parameter $\zeta$ is shown.
A significantly stronger increase of $\kappa$ with increasing $r$
was predicted in Ref.~\cite{barak:1}. The difference between the
results presented here and those obtained in Ref.~\cite{barak:1}
stem from the sign error pointed out after Eq.~(\ref{eq:app_out}).

\begin{figure}
\unitlength1cm
\begin{picture}(18,7)(0,0)
\put(0,0) {\includegraphics[width=10.cm]{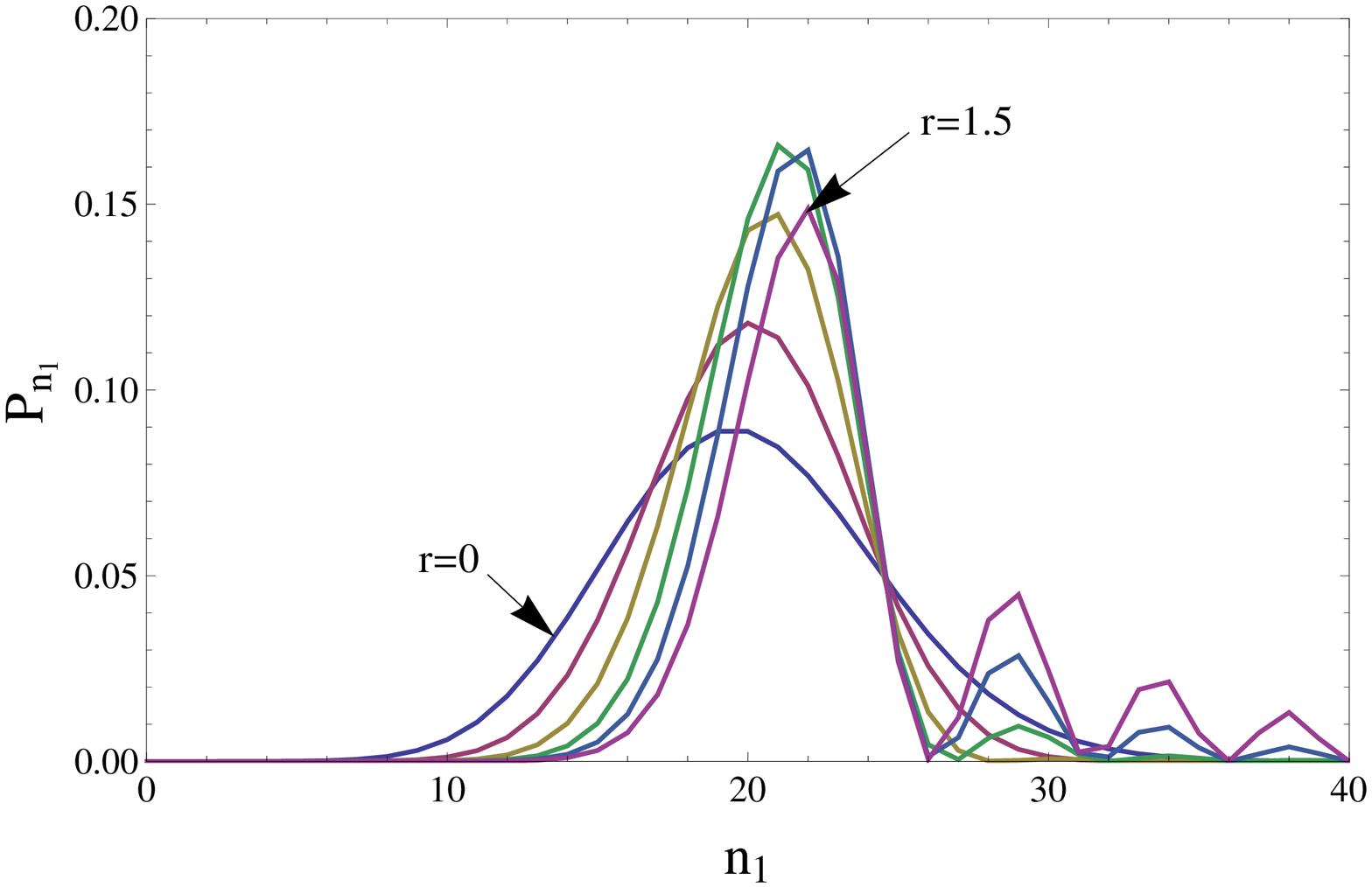}}
\put(10.1,3.5){\includegraphics[width=4. cm]{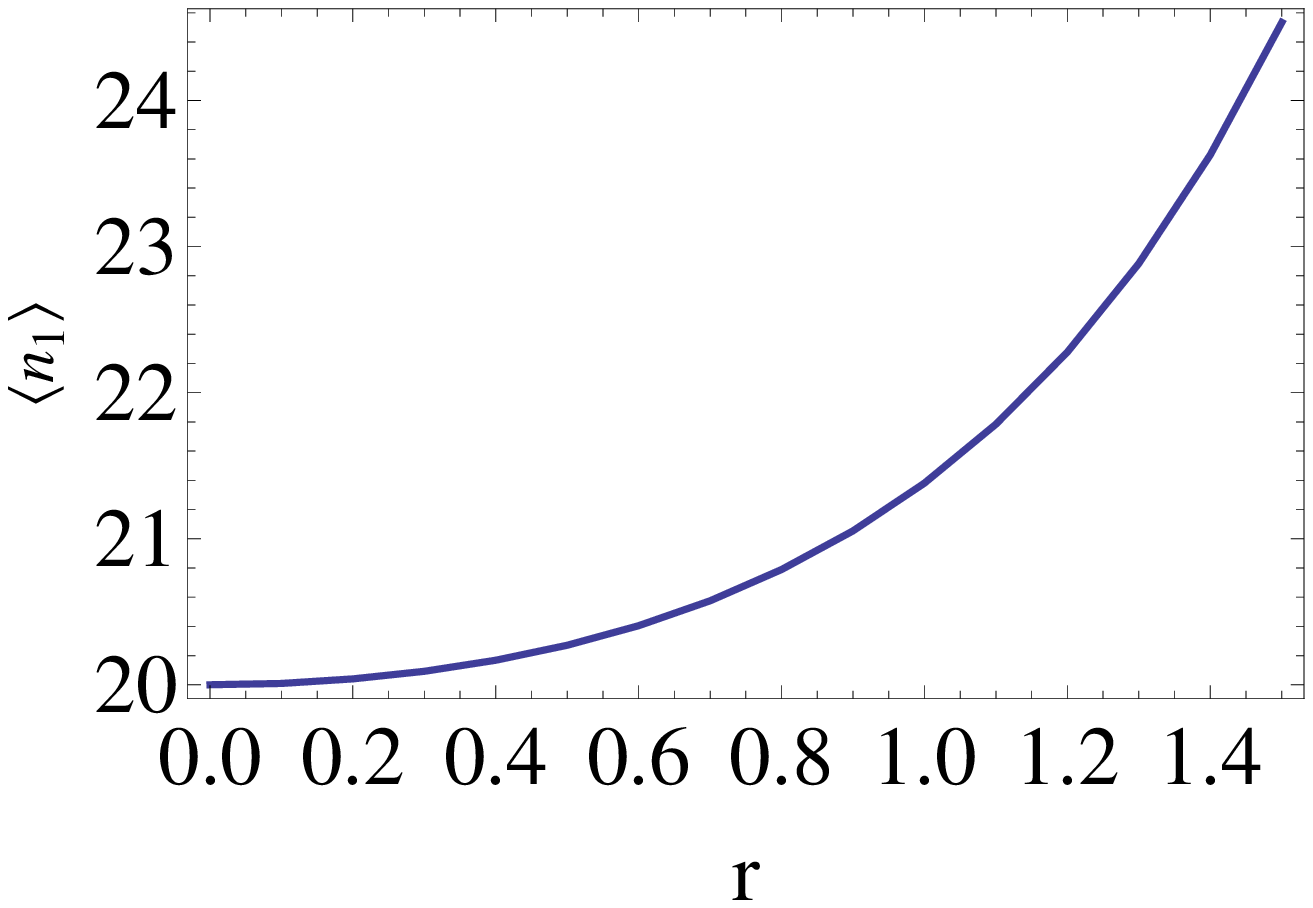}}
\put(10.0,.5){ \includegraphics[width=4. cm]{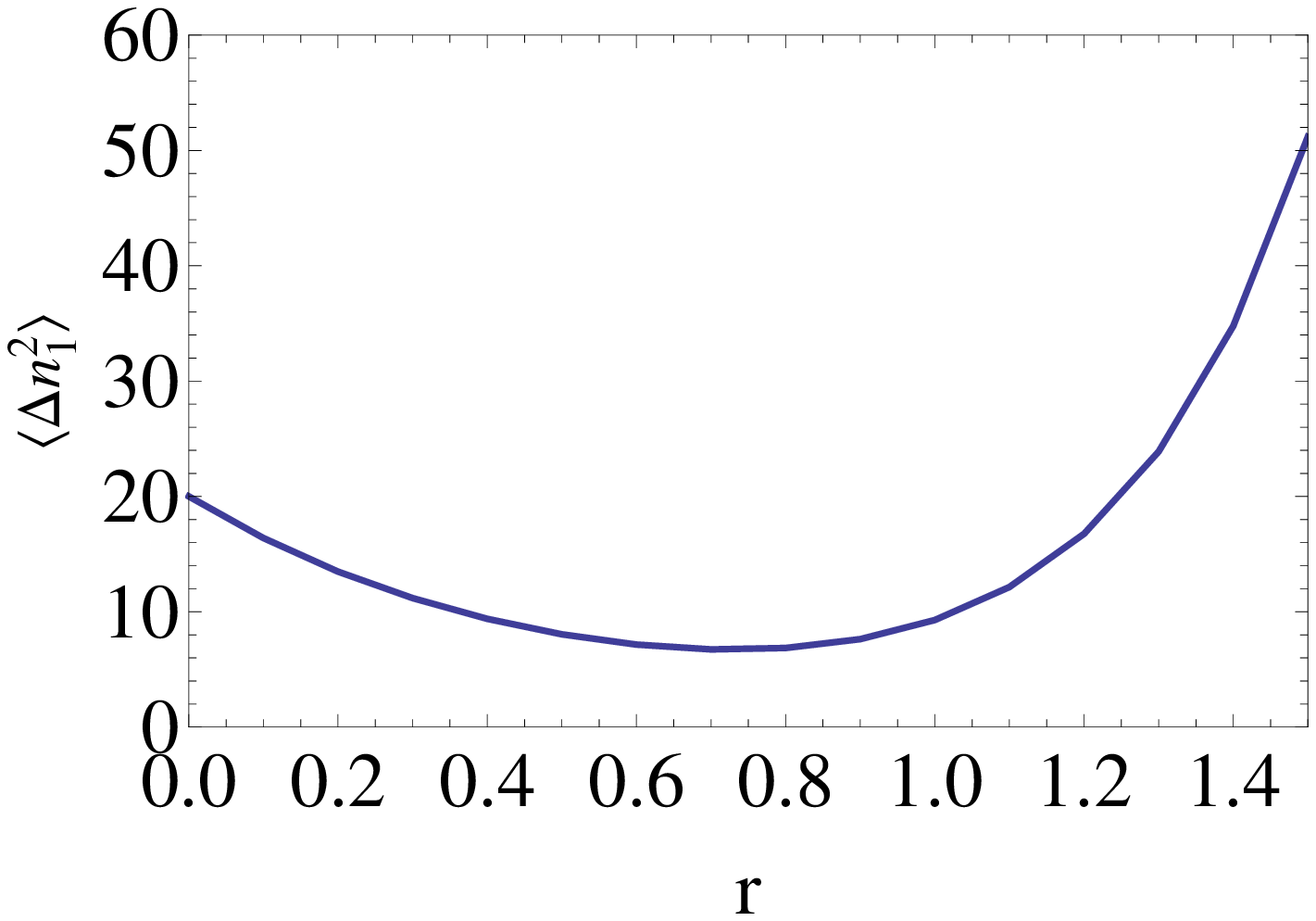}}
\end{picture}
\caption{(Color online) Photon number probability distribution for $|\alpha\delta|^2=20$, $\theta=2\phi$,
and different values of the squeezing parameter $r=0,~0.3,~0.6,~0.9,~1.2,~1.5$ in the dark port.
On the right hand side of the plot we show the mean and variance squared of these distributions.}
\label{fig:phot_distr10}
\end{figure}

\begin{figure}
\unitlength1cm
\begin{picture}(18,7)(0,0)
\put(0,0)     {\includegraphics[width=10.cm]{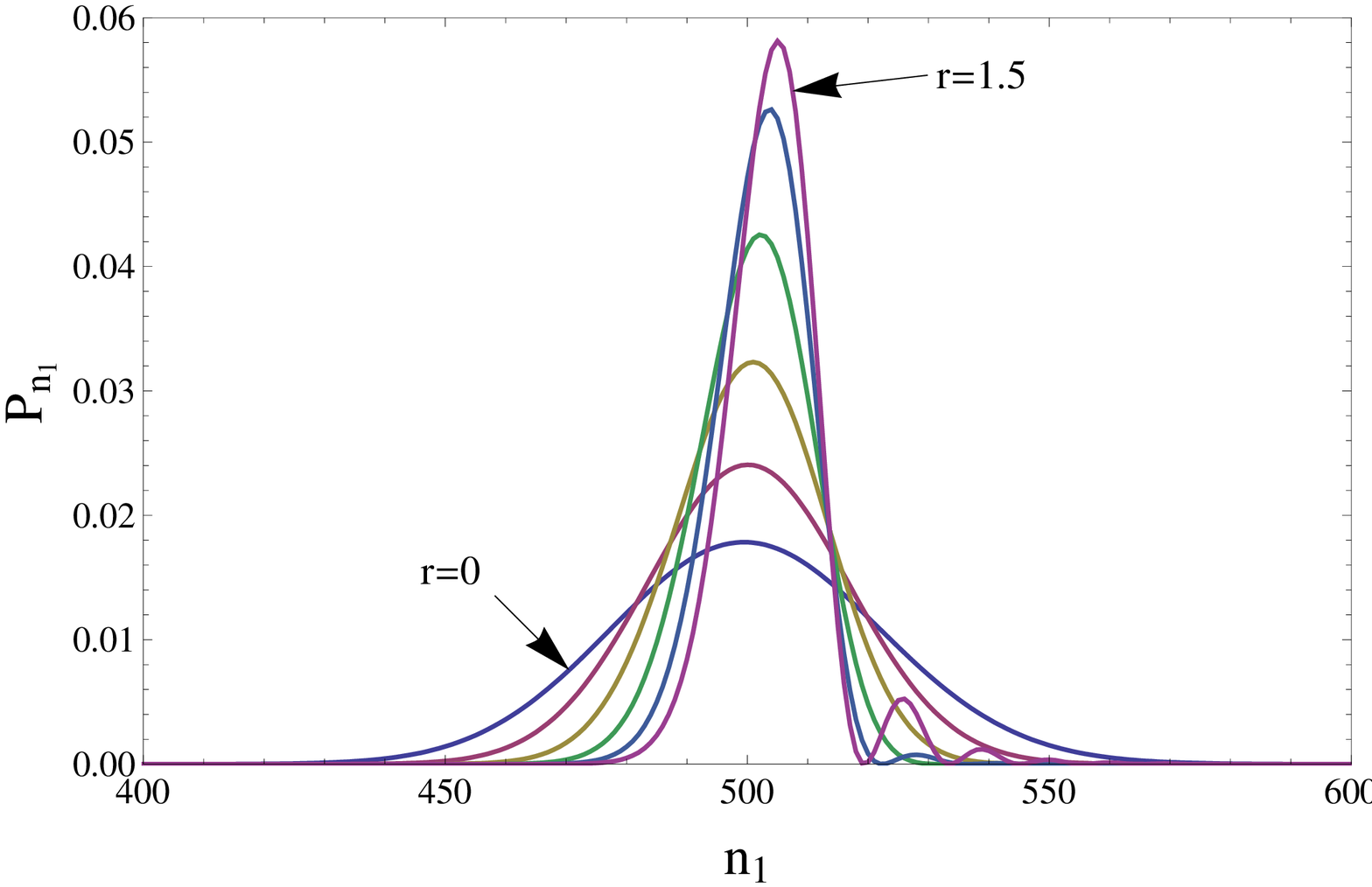}}
\put(10.1,3.5){\includegraphics[width=4.cm]{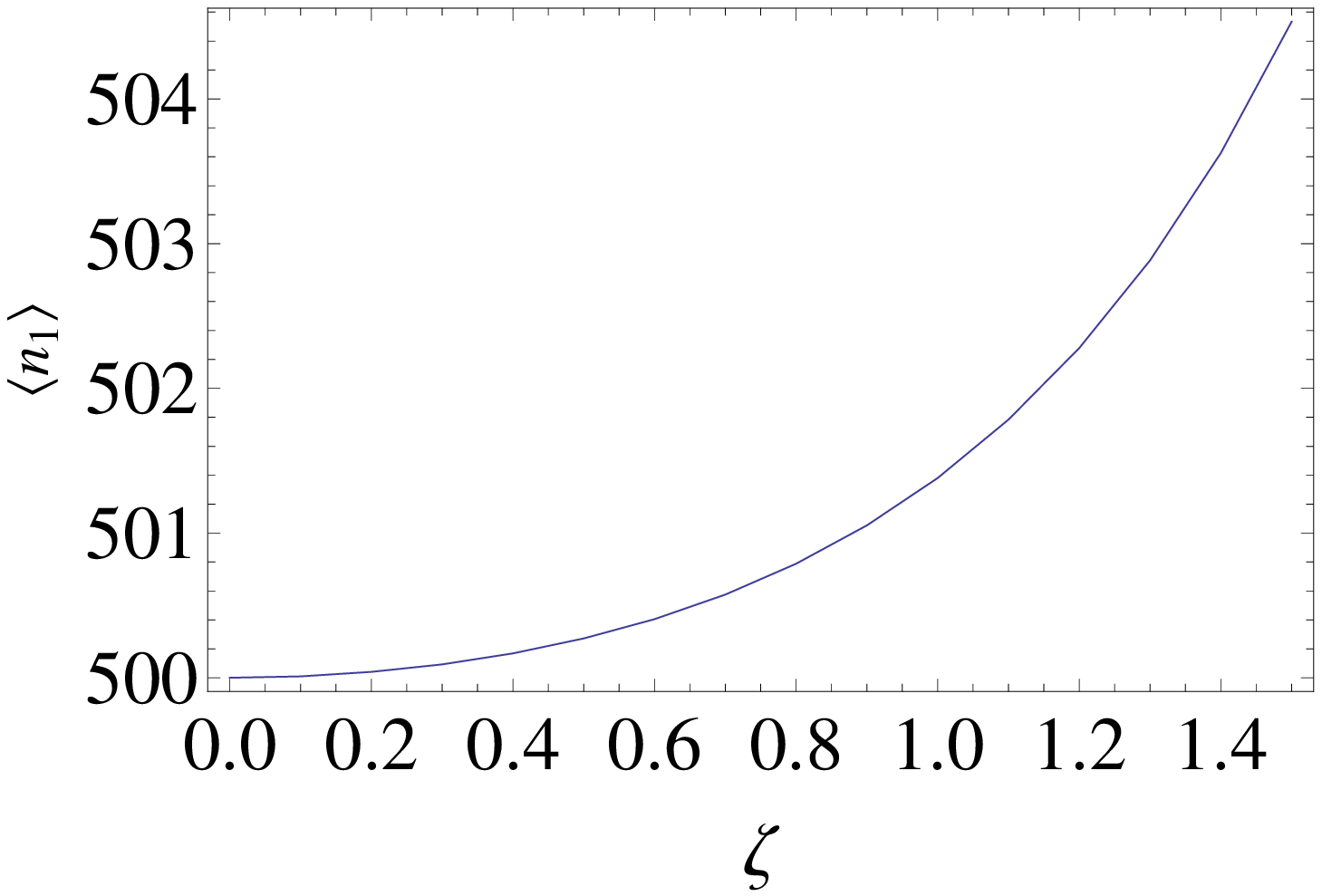}}
\put(10,.5)   { \includegraphics[width=4.cm]{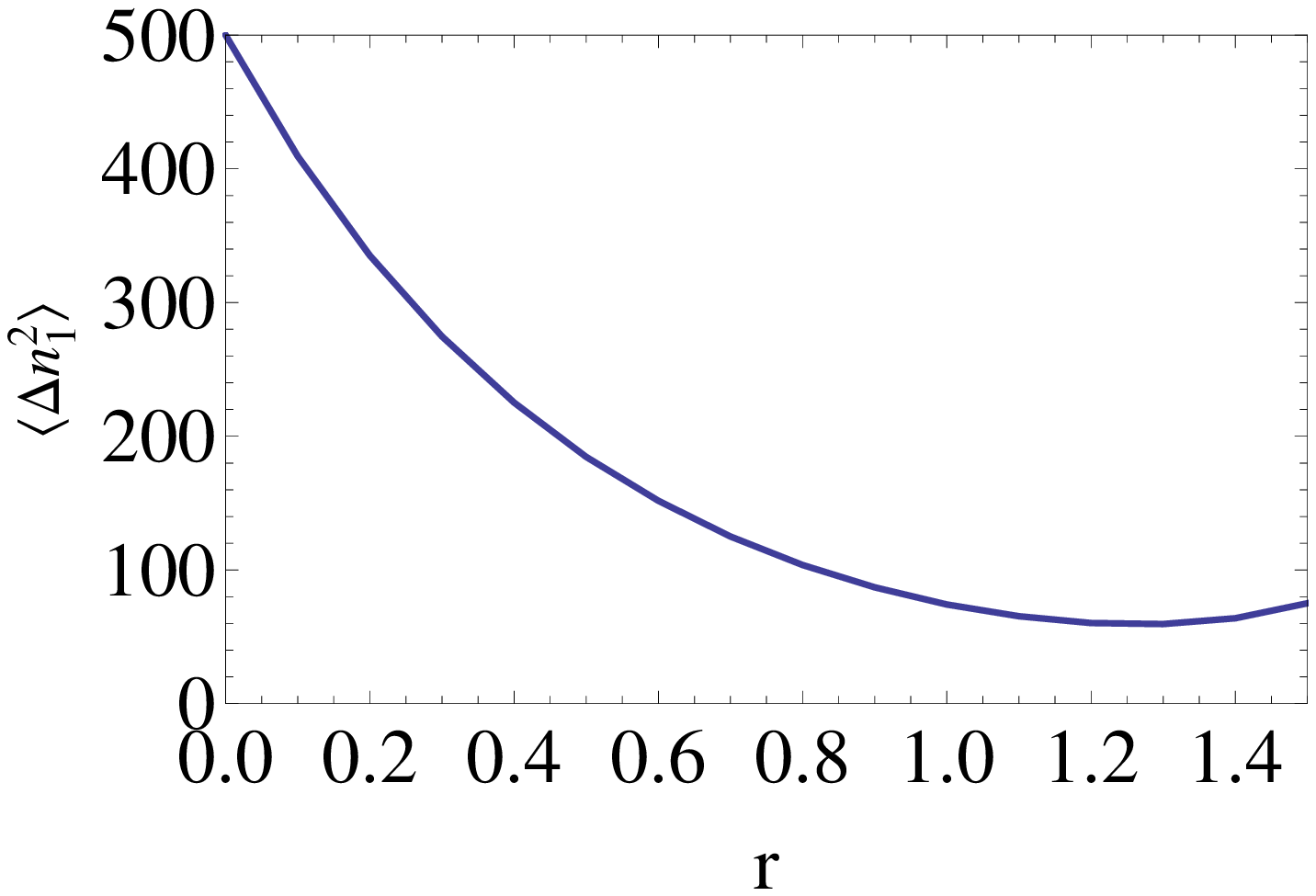}}
\end{picture}
\caption{(Color online) Photon number probability distribution for $|\alpha\delta|^2=500$, $\theta=2\phi$, and different
values of the squeezing parameter $r=0,~0.3,~0.6,~0.9,~1.2,~1.5$ in the dark port.
On the right hand side of the plot we show the mean and the variance squared of these distributions.
}\label{fig:phot_distr500}
\end{figure}

With $\kappa$ as input we calculate the photon number distribution
for the output state from Eqs.~(\ref{eq:darkdist}) for
$|\delta\alpha|^2=20$ and $|\delta\alpha|^2=500$, respectively.
Furthermore, we determine  the mean as well as the variance of these
distributions from Eq.~(\ref{eq:meanvari}). Results of these
calculations are shown in Figs.~\ref{fig:phot_distr10} and
\ref{fig:phot_distr500}.

Optimally we choose a squeezing parameter $\zeta$ and a coherence
parameter $\alpha$ so as to amplify the signal in the output port 1
without increasing its noise. Unfortunately, this does not seem to
be possible in contrast to the conclusions reached in
Ref.~\cite{barak:1}. Analysis of Eqs.~(\ref{eq:meanvari}) as a
function of $r$ and $\theta-2\phi$ indicates that a significant
amplification of the signal in output port 1 is not possible within
the parameter ranges experimentally accessible and considered here.
In particular, for $\theta=2\phi$ one finds $\lambda=0$ and $\langle
n_1 \rangle =\delta^2|\alpha|^2 + \sinh^2 r$.

\begin{figure}
\unitlength1cm
 \center  {\includegraphics[width=6.5cm]{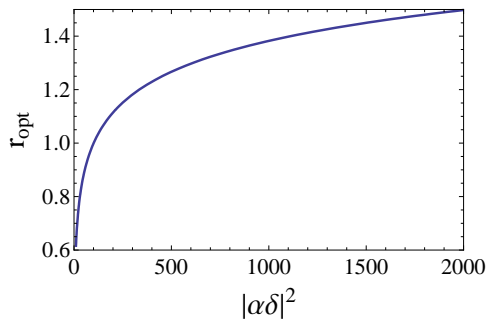}}
\caption{(Color online) Optimal squeezing parameter as a function of $|\alpha\delta|^2$
determined from the minimum of the variance of the distribution function.}\label{fig:opt}
\end{figure}

Only the noise can be slightly influenced: a numerical study of the
variance of the photon distribution given in
Eqs.~(\ref{eq:meanvari}) shows that  $\theta=2\phi$ and $\lambda=0$
is the optimal choice. Then, one finds $\langle \Delta n_1^2 \rangle
= \delta^2|\alpha|^2 \e^{-2r} +2 \sinh^2 r \cosh^2r$. The minimum of
the variance depends on the parameter $\alpha\delta$, and in
Fig.~\ref{fig:opt} we show the optimal squeezing parameter as a
function of $|\delta\alpha|^2$. This squeezing parameter should be
chosen in order to minimize photon counting uncertainties. For large
squeezing parameters the distributions show characteristic
oscillations~\cite{scully:1}.

\section{Summary and conclusions}

In a recent paper by Barak and Ben-Aryeh~\cite{barak:1} it was
suggested that due to entanglement effects in the output state of a
beam splitter, the gravitational wave signal observed with an
interferometer may be amplified without simultaneously increasing
quantum mechanical photon counting noise.

In this paper we reanalyze such entanglement effects, and cannot
confirm the significant signal amplification predicted in
Ref.~\cite{barak:1} but agree that consideration of such
entanglement effects is important in order to calculate the photon
counting noise in optical interferometers.  We observe that for a
strong coherent input in port 1 of the beam splitter and a squeezed
vacuum injected in port 2 photon counting statistics slightly
improves if the beam splitter is oriented slight off the right angle
with the incoming strong coherent field. The differences between the
calculations presented here and those of Ref.~\cite{barak:1} can be
traced to algebraic issues, which are discussed explicitly in
section~\ref{se:numres}.

It appears that entanglement effects must be carefully studied for
any particular interferometer design, a task for which the formulas
developed in the present paper may be helpful, in particular the
general result Eq.~(\ref{eq:phodi}) for the calculation of photon
distributions in the output state.

\appendix

\section{Disentangling}\label{se:disentangling}

In order to solve the disentangling problem ($r=|\zeta|$)
\begin{equation}\label{eq:disenprob}
\exp(r\hat{A})= \exp(\sigma_{T} \hat{t}_{12})\exp(\sigma_{S} \hat{s}_{12})\exp(\sigma_1 \hat{s}_1)\exp(\sigma_2 \hat{s}_2)
\end{equation}
with $\hat{A}$ given in Eq.~(\ref{eq:A}), we consider the Lie
algebra of the operators $\left\{\hat{s}_1, \hat{s}_2, \hat{s}_{12},
\hat{t}_{12}\right\}$ defined in Eqs.~(\ref{eq:oper1}) and
(\ref{eq:oper2}) with their commutators given in
Eq.~(\ref{eq:commutators}). The corresponding matrix representations
of these operators
\begin{equation}
\hat{s}_1 =
\left(
\begin{array}{ll}
1 & 0 \hbox{} \\
0 & 0 \hbox{}
\end{array}
\right),\quad
\hat{s}_2= \left(
\begin{array}{ll}
0 & 0 \hbox{} \\
0 & 1 \hbox{}
\end{array}
\right),\quad
\hat{s}_{12}= \left(
\begin{array}{ll}
0 & -1 \hbox{} \\
-1 & 0 \hbox{}
\end{array}
\right),\quad
\hat{t}_{12}= \left(
\begin{array}{ll}
0 & -1 \hbox{} \\
1 & 0 \hbox{}
\end{array}
\right)
\end{equation}
fulfill the same commutation relations.

The matrix representation of the operator $\hat{A}$ in
Eq.~(\ref{eq:disenprob}) is then obtained as
\begin{equation}
A=\left(
\begin{array}{ll}
\sin^2\gamma & -\cos\gamma\sin\gamma \hbox{} \\
-\cos\gamma\sin\gamma & \cos^2\gamma \hbox{}
\end{array}
\right).
\end{equation}
Since $\hat{A}^n=\hat{A}$ for all $n \in N$ we find for the left
hand side of Eq.~(\ref{eq:disenprob})
\begin{equation}\label{eq:left}
\exp(r\hat{A})=\sum_{n=0}^{\infty}{\frac{(r\hat{A})^n}{n!}}=\hat{I} + (\e^{r}-1)\hat{A}.
\end{equation}

The matrix representation of the right hand side
Eq.~(\ref{eq:disenprob}) is also easily calculated
\begin{eqnarray}\label{eq:right}
& &\exp(\sigma_{T} T_{12})\exp(\sigma_{S} \hat{s}_{12})\exp(\sigma_1 \hat{s}_1)\exp(\sigma_2 \hat{s}_2)= \\
& &
\left(
\begin{array}{ll}
\e^{\sigma_1}(\sin\sigma_{T}\sinh\sigma_{S}+\cos\sigma_{T}\cosh\sigma_{S}) & -\e^{\sigma_2}(\sin\sigma_{T}\cosh\sigma_{S}+\cos\sigma_{T}\sinh\sigma_{S}) \hbox{} \\
\e^{\sigma_1}(\sin\sigma_{T}\cosh\sigma_{S}-\cos\sigma_{T}\sinh\sigma_{S}) & -\e^{\sigma_2}(\sin\sigma_{T}\sinh\sigma_{S}-\cos\sigma_{T}\cosh\sigma_{S}) \hbox{}
\end{array}
\right).\nonumber
\end{eqnarray}

By equating Eqs.~(\ref{eq:left}) and (\ref{eq:right}) one obtains
four equations for the four parameters ${\sigma_T, \sigma_S,
\sigma_1, \sigma_2}$ to be determined
\begin{eqnarray}\label{eq:nonlinear}
\e^{r}\sin^2\gamma + \cos^2\gamma &=& \e^{\sigma_1}(\sin\sigma_T\sinh\sigma_S+\cos\sigma_T\cosh\sigma_S),\nonumber\\
(1-\e^{r})\cos\gamma\sin\gamma &=& -\e^{\sigma_2}(\sin\sigma_T\cosh\sigma_S+\cos\sigma_T\sinh\sigma_S),\\
(1-\e^{r})\cos\gamma\sin\gamma&=&\e^{\sigma_1}(\sin\sigma_T\cosh\sigma_S-\cos\sigma_T\sinh\sigma_S),\nonumber\\
\e^{r}\cos^2\gamma + \sin^2\gamma &=&-\e^{\sigma_2}(\sin\sigma_T\sinh\sigma_S-\cos\sigma_T\cosh\sigma_S).\nonumber
\end{eqnarray}
This set of equations is easily solved numerically. Results are
shown in Fig.~\ref{fig:discoeff}.

For $\gamma=\pi/2$  solutions of Eqs.~(\ref{eq:nonlinear}) can
easily be found analytically,
\begin{equation}
\gamma=\pi/2,\quad \sigma_1=r,\quad \sigma_2=0,\quad \sigma_S=0,\quad \sigma_T=0.
\end{equation}
If one expands the trigonometric functions on the right hand side of
Eqs.~(\ref{eq:nonlinear}) keeping only linear terms in the
disentangling coefficients, one finds
\begin{equation}
\gamma=\pi/2+\delta,\quad \sigma_1=r,\quad \sigma_2=0,\quad \sigma_S=-\delta\sinh r,\quad\sigma_T=\delta(1-\cosh r).
\end{equation}
For  $\gamma=\pi/4$ one finds
\begin{equation}
\gamma=\pi/4,\quad \sigma_1=r/2,\quad \sigma_2=r/2,\quad \sigma_S=r/2, \quad\sigma_T=0.
\end{equation}

\section{Evaluation of the entanglement factors}\label{se:eval}
The entanglement factors $\exp(\sigma_T \hat{t}_{12})$ and
$\exp(\sigma_S \hat{s}_{12})$ must be further transformed for
convenient calculations. Using the fact that the operators
$\hat{b}_1^{\dagger}\hat{b}_2^{\dagger}$,   $\hat{b}_1\hat{b}_2$,
and $\frac{1}{2}(1+\hat{b}_1^{\dagger}\hat{b}_1+
\hat{b}_2^{\dagger}\hat{b}_2 )$  form an SU(1,1) Lie algebra one
easily finds that
\begin{equation}
\label{disent_S}
\exp[\sigma_S \hat{s}_{12}] = \exp\left(\hat{b}_1^{\dagger}\hat{b}_2^{\dagger} \e^{i\theta}\tanh \sigma_S \right)
                              \exp\left(- \left(1+\hat{n}_1 + \hat{n}_2 \right)\ln (\cosh \sigma_S )  \right)
                              \exp\left(-\hat{b}_1\hat{b}_2 \e^{- i\theta}\tanh \sigma_S \right).
\end{equation}
as was proved e.g. in Ref.~\cite{dasgupta:1}. Similarly one proves
that
\begin{equation}
\exp[\sigma_T \hat{t}_{12}] = \exp\left(\hat{b}_1\hat{b}_2^{\dagger} \e^{i\theta}\tan \sigma_T \right)
                              \exp\left(-\left(\hat{n}_1 - \hat{n}_2 \right)\ln (\cos \sigma_T ) \right)
                              \exp\left(-\hat{b}_1^{\dagger}\hat{b}_2 \e^{- i\theta}\tan \sigma_T \right).
\end{equation}
using the fact that the operators $\hat{b}_1\hat{b}_2^{\dagger}$,
$\hat{b}_1^{\dagger}\hat{b}_2$, and
$-\frac{1}{2}\left(\hat{b}_1^{\dagger}\hat{b}_1-
\hat{b}_2^{\dagger}\hat{b}_2 \right)$  form an SU(2) Lie algebra.
Expanding out the exponentials in the expressions above allows for
convenient application of these operators on Fock states.

\section{Transformations}\label{se:transformations}
The following transformations for products of displacement operators
and squeezing operators can be easily proved from the
Baker-Campbell-Haussdorff theorem and the Boson commutation
relations,
\begin{eqnarray}
\hat{D}(\alpha_2)\hat{D}(\alpha_1)&=&\hat{D}(\alpha_1+\alpha_2) \exp \left[\frac{1}{2}(\alpha_2 \alpha_1^* - \alpha_2^*\alpha_1) \right],\nonumber\\
\hat{D}(\alpha)\hat{S}(\zeta)&=&\hat{S}(\zeta)\hat{D}(\alpha \cosh r + \alpha^* \e^{i \theta}\sinh r), \quad\zeta=r \e^{i\theta}.
\end{eqnarray}
These relations are used to obtain Eq.~(\ref{eq:outpihalf}).

\vskip 2cm

{\bf Acknowledgements}

We would like to thank Yacob Ben-Aryeh for a useful correspondence
and a critical reading of the manuscript. M. W. acknowledges Roman
Schnabel for a useful correspondence and Robert Wynands for useful
discussions on experimental issues. V.G.V. thanks
Physikalisch-Technische Bundesanstalt for support of an internship,
while this work was performed.


\end{document}